# The different responses of universities to introduction of performance-based research funding

Giovanni Abramo*

*Laboratory for Studies in Research Evaluation*
*at the Institute for System Analysis and Computer Science (IASI-CNR)*
*National Research Council of Italy*
ADDRESS: Istituto di Analisi dei Sistemi e Informatica, Consiglio Nazionale delle Ricerche, Via dei Taurini 19, 00185 Roma - ITALY
ORCID: 0000-0003-0731-3635 - giovanni.abramo@uniroma2.it

Ciriaco Andrea D’Angelo

*University of Rome ‘Tor Vergata’ - Italy and*
*Laboratory for Studies in Research Evaluation (IASI-CNR)*
ADDRESS: Dipartimento di Ingegneria dell’Impresa, Università degli Studi di Roma ‘Tor Vergata’, Via del Politecnico 1, 00133 Roma - ITALY
ORCID: 0000-0002-6977-6611 - dangelo@dii.uniroma2.it

## Abstract

Governments and organizations design performance-based research funding systems (PBFRS) for strategic aims, such as to selectively allocate scarce resources and stimulate research efficiency. In this work we analyze the relative change in research productivity of Italian universities after the introduction of such a system, featuring financial and reputational incentives. Using a bibliometric approach, we compare the relative research performance of universities before and after introduction of PBFRS, at the overall, discipline and field levels. The findings show convergence in the universities’ performance, due above all to the remarkable improvement of the lowest performers. Geographically, the universities of the south (versus central and northern Italy) achieved the greatest improvement in relative performance. The methodology, and results, should be of use to university management and policy-makers.

## 1. Introduction

Never has there been such deep and widespread awareness of the importance of scientific research among populations, as in this dramatic period of the Covid-19 pandemic, even though for decades, far-sighted governments have identified the strengthening of technical-scientific infrastructure as a top policy priority, and within this, the strengthening of higher education.

Country after county has moved to strengthen competitive mechanisms in the research sector, particularly through merit-based access to resources. Numerous countries have introduced performance-based research funding systems (PBRFS), intended to stimulate continuous improvement of research effectiveness and efficiency (Hicks, 2012; Lewis, 2013; Woelert, 2015; Geuna & Piolatto, 2016). Among the EU28 (including the UK), 16 countries have implemented some form of PBRFS, and there have been attempts at analysis of the strengths and weaknesses of the different approaches (Zacharewicz, Lepori, Reale, & Jonkers, 2019).

PBRFS are generally associated with national research assessment exercises, whereby public resources are then allocated to the research organizations on the basis of the assessments. More rarely, the national research assessments have been conducted without directly links to funding. In addition to fostering higher productivity and channeling resources to the highest-performing organizations, the results of these exercises contribute to strategic analyses, in turn informing national research policies, and to reducing information asymmetry on the ‘knowledge market’, between supply (research organizations) and demand (students, industry, etc.), thereby gaining efficiency. Not least, the assessment exercises also serve to demonstrate to tax-payers that investment in research is effective, and delivers public benefits.

For the research organizations, the incentive to improve arrives not only from the financial leverage associated with evaluation exercises (OECD, 2010; Woelert & McKenzie, 2018), but also from the goodwill earned by ranking in the top positions. The return of image is reflected in the increased attraction of talented researchers, professors, students, private funding, donations, etc.

Not all institutions under evaluation respond in the same way to the PBRFS incentives. The response depends on several factors, most important the degree to which national incentives have been deployed within the individual research organizations, but also the researchers’ sense of identification and belonging to their own organization. A number of studies have attempted to assess the impact of national assessment exercises on the research behavior and performance of individuals and organizations. This study inserts in this stream of research. In particular, we aim to verify whether and to what extent Italian universities have improved their relative research performance (i.e. rank, not score), as a response to the PBRFS introduced in association with the VQR 2004-2010 research assessment exercise, launched in late 2011.

Assessing the variation in relative performance (i.e. compared to other universities) is very different from assessing the variation in absolute performance: a university can, for example, improve its relative performance even if its absolute performance decreases (i.e. on average, other universities suffered greater deterioration).

Assessing relative performance is also much easier. Instead, determining a causal effect of PBRFS on absolute research performance is a formidable task, because of the difficulty of controlling for all possible confounding variables. For example, if not controlled for, the constant increase in number of journals, indexed by the bibliographic repertories at the basis of the bibliometric assessments, can bias the analytical results. This factor is largely irrelevant in the case of assessing relative change in performance, when any before-after contextual changes with potential impact on performance would only be problematic in the event they have systematic pro/con effects on particular groups of universities. Among the latter, there could occur changes in the resources allocated to research by each single university. This would be practically impossible to trace, especially at individual/field level, as a correct assessment of performance requires.

Notwithstanding a number of assumptions and limitations common to bibliometric studies, our study aims to improve the existing literature in several ways: i) we assess performance, by the same indicator, before and after the introduction of the PBRFS; ii) we adopt a performance indicator derived from the micro-economic theory of production, which integrates the quantity and quality dimensions of performance, accounts for individual contributions to output, and above all inserts controls for inputs, where such data are available; iii) to assess performance, we measure all scientific output indexed in the Web of Science (WoS) core collection, and not just the small share of products submitted by universities to the VQR; and iv) we exploit a fine-grained field classification of Italian professors, which allows control for the different intensity of publications across fields, and avoids comparing apples to oranges (Abramo, Cicero, & D'Angelo, 2013a).

In assessing the response of individual universities to PBRFS we are also able to assess whether the variability of performance between universities has changed (convergence vs dispersion), and if there has been any reduction in the regional divide in universities, a sensitive topic in a number of countries and very much so in Italy, for the so-called north-south divide (Checchi, Mazzotta, Momigliano, & Olivanti, 2020; Grisorio & Prota, 2020; Abramo, D'Angelo, & Rosati, 2016; Viesti, 2016).

The study is articulated as follows. In the next section we review the literature on the subject. In Section 3 we present the Italian VQR. In Section 4 we illustrate the method and field of analysis. In Section 5 we present the results. Section 6 concludes the work with our remarks and policy recommendations.

## 2. Literature review

Research evaluations affect motivation, performance, individual and organizational behavior. They enter into policies on recruitment and career advancement, diversity, disparity, gender differences and academic freedom, as well as on the relative importance of teaching and research, and influence technology transfer and the arrangement of loci of power. Some reviewers of PBRFS state that there is still not sufficient evidence on the effects of these schemes, and that our understanding of the proper design and implementation of these schemes is still incomplete (de Boer et al., 2015). The extant literature on the impact of PBRFS can be classified into three somewhat overlapping streams of research. A first line concerns the impact on the strategic and organizational management of research institutions, and another, the effects of PBRFS on individual behavior, especially the perverse effects. Our own study contributes to a third line of research, concerning the macro- or micro-economic impacts of PBRFS in terms of increases in research performance, convergence or divergence within and between institutions, and regions.

For scholars advancing the first line of research, the central question is how organizations respond to PBRFS. They investigate to what extent system-wide incentives can affect local management practices, and how these in turn might improve the organizational effectiveness, efficiency, and merit-based resource management practices (Espeland & Sauder, 2007; Sauder & Espeland, 2009). A number of studies have analyzed the deployment of the incentive schemes embedded in national PBRFS within individual research organizations. In Australia, it was found that most universities tend to replicate the major national performance indicators internally (Woelert & McKenzie, 2018). In Norway, mixed practices resulted: while a substantial replication was found in a number of institutions, large variation across organizations, disciplines and departments was also observed (Aagaard, 2015).

Other scholars have concentrated on the effects of PBRFS on individual behavior, which are especially likely when the organizations deploy the national incentives at individual level (Moher et al., 2018). Most studies have investigated the unintended effects (De Rijcke, Wouters, Rushforth, Franssen, & Hammarfelt, 2016), among others: i) scientific misconduct, such as 'salami slicing', plagiarism, self-plagiarism, inappropriate self-citations, scientific fraud (Hazelkorn, 2010; Edwards & Roy, 2017; Seeber, Cattaneo, Meoli, & Malighetti, 2019); ii) discouragement of research diversification, interdisciplinary and innovative research (Hicks, 2012; Rafols, Leydesdorff, O'Hare, Nightingale, & Stirling, 2012; Wilsdon, 2015; Abramo, D'Angelo, & Di Costa, 2018); and iii) the trade-off between research and teaching (Enders, Kehm, & Schimank, 2015; De Philippis, 2015).

The third stream of research concerns the PBRFS impact on overall performance of institutions under evaluation. Most often, the impact on performance is analyzed along two separate dimensions: number of articles and average impact. Evidence is rather mixed, not only across countries, but also from different investigators within the same country (Adams & Gurney, 2010; Wang & Hicks, 2013). In some cases, changes are analyzed along the performance criteria specifically embedded in the quality assessment exercises, comparing performance scores before and after (Buckle, Creedy, & Gemmell, 2020; Checchi, Mazzotta, Momigliano, & Olivanti, 2020).

A number of studies have been designed to respond to governments' needs to assess the impact of their PBRFS on national research productivity. Several have assessed the before-after variations in the numbers of publications and average citations per paper (UK: McNay, 2015; Australia: Butler, 2003a, 2003b; van den Besselaar, Heyman, and Sandström, 2017; Norway: Schneider, Aagaard, & Bloch, 2016). A recent investigation of 18 countries with some form of PBRFS in operation reported a positive effect on average impact per paper, positive but temporary effect on number of publications, and almost unvaried shares of articles published in high IF-journals (Checchi, Malgarini, & Sarlo, 2019). Cattaneo, Meoli, and Signori (2016) investigated the change in publications per faculty member following the introduction of the first Italian research assessment exercise in 2003, focusing on the moderating effect of university "legitimacy", defined as their "level of recognition based on the adherence to socially accepted norms and expectations". Other studies have assessed the variations in institutions' performance scores resulting from the quantitative evaluation criteria of the national exercises (New Zealand: Buckle & Creedy, 2019; Italy: Checchi, Mazzotta, Momigliano, & Olivanti, 2020).

In addition to the inevitable obstacle of controlling for all confounding factors, the main fault of the above studies is the selection of the performance indicator. Neither indicator - number of publications or average impact per paper - are valid indicators of performance when considered alone. When both indicators are measured, but the variations are divergent, the performance question remains unresolved. The failure to adopt fractional counting of publications leads to error too (Waltman & van Eck, 2015).[1] Furthermore, when evaluation exercises are based on a small share of the research organization products, performance scores are highly distorted (Abramo, Cicero, & D'Angelo, 2013b), and can depend on the (in)ability of the research organizations to select their best products (Abramo, D'Angelo, & Di Costa, 2014). In our view, there is little sense in comparing such distorted performance scores, as done by Checchi, Mazzotta, Momigliano, and Olivanti (2020) for the Italian case. Moreover, any variations in performance scores might be due to a better or worse selection of submitted products.

When interpreting or comparing the results of any investigations on impact of PBRFS on research institutions, particular attention to the choices made in terms of performance indicators and assessment methods should be addressed.

## 3. The 2004-2010 VQR Italian research assessment exercise

In 2010 the Italian Ministry of University and Research (MIUR) established the Agency for Evaluation of University and Research Systems (ANVUR), entrusted with implementation of national research assessment exercises, whose results were to guide the merit-based allocation of funds to universities and research institutions under ministry responsibility. ANVUR opened the "Evaluation of Research Quality" or "VQR 2004-2010" process in November 2011, and brought it to completion in July 2013 with publication of the university performance ranking lists.[2] The first VQR marked a turning point in the funding policies for Italian universities, for the first time allocating a significant share of core resources, known as Ordinary Finance Funds (FFO), in function of the assessment: 13% in 2013, increasing stepwise to now 25%.

The first VQR came at a moment of particularly strong incentive to achieve high rank, as the severe national financial constraints of previous years had led to heavy cuts to universities. VQR 2004-2010 was followed by VQR 2011-2014, and now, VQR 2015-2019.[3]

The subjects under evaluation are the institutions, their macro-disciplinary areas and departments, but not the individual researchers. The VQR 2004-2010 determined the evaluation of the overall institutions by the weighted sum of a number of indicators: 50% from a score for the quality of the research products submitted, and 50% from a composite of six other indicators (10% each for ability to attract research funds, mobility of research staff, internationalization, and PhD programs; 5% each for allocation of internal funds to research, and for improvement with respect to previous performance).

As in other PBRFS, while inputs are individual, rewards are collective. The efforts from individual academics may be enormous, and their participation in the evaluation process is certainly onerous, yet the financial recognition goes to the institutions, which can then deploy resources as they wish.

ANVUR appointed 14 evaluation panels (GEVs)[4] of national and foreign experts, one for each university disciplinary area (UDA) in the national academic system. The institutions subject to evaluation were to submit a specific number of products for each person on their research staff, in function of academic rank and their period of activity over the seven years considered. The university faculties required their individual members to submit up to three products.

---

[1] The authors argue that "properly field-normalized results cannot be obtained when full counting is used", and recommend "the use of fractional counting in bibliometric studies that require field normalization, especially in studies at the level of countries and research organizations.

[2] This description of the VQR is similar to that in other publications by the authors.

[3] https://www.anvur.it/attivita/vqr/ last accessed on 24 November 2020.

[4] Acronym of "Groups of Evaluation Experts"

ANVUR defined the acceptable products as: a) journal articles; b) books, book chapters and conference proceedings; c) critical reviews, commentaries, book translations; d) patents; e) prototypes, project plans, software, databases, exhibitions, works of art, compositions and thematic papers.

Any results produced in collaboration with professors in the same institution could only be presented once. Professors were therefore typically asked to identify a set of products larger than the minimal demand, from which the administration could complete the selection of the numbers required for the VQR evaluation. The products were then submitted to the appropriate GEVs based on the professor's identification of the field for each product. The GEVs were to judge the merit of each product as one of four values:

A = Excellent (score 1), if the product places in the top 20% on "a scale of values shared by the international community";
B = Good (score 0.8), if the product places in the 60%-80% range;
C = Acceptable (score 0.5), if the product is in the 50%-60% range;
D = Limited (score 0), if the product is in the bottom 50%.

The institutions were also subject to penalties:

i. in proven cases of plagiarism or fraud (score -2);
ii. for product types not admitted by the GEV, or lack of relevant documentation, or produced outside the 2004-2010 period (score -1);
iii. for failure to submit the requested number of products (-0.5 for each missing product).

This last penalty considered the nature of the Italian higher education system, which unlike the systems of English-language countries does not provide for both "teaching-only" and research universities. In keeping with the Humboldtian university model, emphasizing the unity of teaching and research, Italy does not have teaching-only universities, and all professors are required to carry out both teaching and research.

It should be noted that both the 2004-2010 and subsequent 2011-2014 VQR have received various criticisms concerning the evaluation criteria, performance indicators, distortions in rankings and the resulting allocation of resources (Abramo & D'Angelo, 2017, 2016, 2015, 2011; Abramo, D'Angelo, & Di Costa, 2014a; Baccini & De Nicolao, 2016; Franceschini & Maisano, 2017). There is also the objection that, as with other PBRFS, the outcomes depend on individual performance of professors but the financial recognition goes to the universities, who are not obliged to deploy the money based on individual performance, or even for promoting research.

## 4. Data and method

For purposes of verifying which (and to what extent) Italian universities have improved their relative research performance as a response to the VQR 2004-2010, we adopt a "before-after analysis". Since the VQR 2004-2010 was launched in late 2011, we measure the bibliometric performance of universities over the 2007-2011 five-year period, assume 2012 as an interval year, and then compare to the performance of universities in over the 2013-2017 period. To avoid bias in the comparison, we adopt identical citation time windows, counting citations at 31/12/2013 for the first five-year period, and at 31/12/2019 for the second.

### 4.1 Data

Data on the faculty at each university were extracted from the database on Italian university personnel, maintained by the MIUR.[5] In Italian universities, all professors are classified in one and only one field, named scientific disciplinary sector (SDS), 370 in all. SDSs are grouped into disciplines, named university disciplinary areas (UDAs), 14 in all.[6] For motives of coverage of the bibliographic repertories, the field of observation is limited to the UDAs (10 in all, containing 199 SDSs) that ANVUR itself defines as "bibliometric", and evaluates with bibliometric indicators.

The analysis dataset is made up of two distinct sets of Italian professors, one for each five-year period:

- for 2007-2011, 30,922 professors (assistant, associate and full) with at least 3 years on staff during the five-year period,
- for 2013-2017, 28,046 professors with at least 3 years on staff in the five-year period.

Table 1 indicates their breakdown by UDA.

***Table 1: Dataset of the analysis***

| | | 2007-2011 | | 2013-2017 | |
|---|---|---|---|---|---|
| UDA* | SDSs | Universities | Professors | Universities | Professors |
| 1 - Mathematics and computer science | 10 | 64 | 2823 | 73 | 2537 |
| 2 - Physics | 8 | 64 | 1931 | 69 | 1783 |
| 3 - Chemistry | 11 | 59 | 2588 | 61 | 2393 |
| 4 - Earth sciences | 12 | 47 | 936 | 51 | 863 |
| 5 - Biology | 19 | 64 | 4256 | 73 | 3885 |
| 6 - Medicine | 50 | 62 | 8735 | 67 | 7102 |
| 7 - Agricultural and veterinary sciences | 30 | 55 | 2725 | 55 | 2598 |
| 8 - Civil engineering | 9 | 53 | 1340 | 59 | 1286 |
| 9 - Industrial and information engineering | 42 | 70 | 4557 | 74 | 4572 |
| 10 - Psychology | 8 | 61 | 1031 | 67 | 1027 |
| Total | 199 | 85 | 30922 | 93 | 28046 |

The two datasets are quite similar, however there is a significant reduction in the teaching staff in the second period (-2876 units, 9% of total) and a redistribution of professors over a larger number of universities (93 in 2013-2017 vs 85 in 2007-2011). The contraction affected all UDAs (particularly Medicine, -19%), with the exception of Industrial and information engineering.

[5] http://cercauniversita.cineca.it/php5/docenti/cerca.php, last accessed on 24 November, 2020.

[6] The complete list is accessible on http://attiministeriali.miur.it/UserFiles/115.htm, last accessed on 24 November, 2020.

### 4.2 Measuring research productivity

Aligning with the micro-economic theory of production, we measure professors' performance by their research productivity. Productivity, the quintessential indicator of efficiency in any production system, is commonly defined as the rate of output per unit of input. But because in research systems, publications (output) have different values (impact), and the resources available for research are unequal across individuals and organizations, an appropriate definition of productivity in research systems is: the value of output per euro spent in research.

The operazionalization of the measurement of research productivity requires the adoption of a number of assumptions due to lack of data. The *FSS* (fractional scientific strength) indicator is a proxy measure of research productivity. A thorough description of the *FSS* indicator and the underlying theory can be found in Abramo and D'Angelo (2014).[7]

To measure the yearly average research productivity of every professor, we use the following formula:[8]

$$FSS_P = \frac{1}{\left(\frac{w_r}{2}+k\right)} \cdot \frac{1}{t}\sum_{i=1}^{N} c_i f_i \qquad [1]$$

where:

$w_r$ = average yearly salary of professor[9]

$k$ = average yearly capital available for research to professor[10]

$t$ = number of years of work by the professor in period under observation

$N$ = number of publications by the professor in period under observation

$c_i$ = weighted combination of field-normalized citations received by publication *i* and field-normalized *IF* of the hosting journal.[11]

$f_i$ = fractional contribution of professor to publication *i*.

The fractional contribution equals the inverse of the number of authors in those fields where the practice is to place the authors in simple alphabetical order but assumes different weights in other cases. For Biology, Biomedical research and Clinical medicine, widespread practice in both Italy is for the authors to indicate the various contributions to the published research by the order of the names in the bylines. For the above disciplines, we thus give different weights to each co-author according to their position in the list of authors and the character of the co-authorship (intra-mural or extra-mural).[12]

---

[7] This description of the measurement of research performance is similar to that in other publications by the authors.

[8] The underlying assumption is that labor and capital contribute equally to production.

[9] We halve labor costs, assuming that 50 per cent of professors' time is allocated to activities other than research.

[10] Sources of input data, and assumptions adopted in the measurement, are found in Abramo, Aksnes, and D'Angelo (2020).

[11] Citations are normalized to the average of distribution of citations received for all WoS-cited publications in same year and SC of publication *i*. IF refers to the year of publication and is normalized to the average of distribution of *IFs* of journals in same SC of publication *i*.

[12] If the publication is produced by exclusively intramural collaboration (only one affiliation in the address list), 40% is attributed to both first and last author, and the remaining 20% is divided among all other authors. In contrast, if the publication address list shows extramural collaborations, 30% is attributed to

The bibliometric dataset used to measure *FSS* is extracted from the Italian Observatory of Public Research, a database developed and maintained by the present authors and derived under license from the Web of Science core collection (WoS).

Data concerning salary costs for research personnel were obtained from the DALIA database,[13] which is also maintained by the MIUR. Professors' salaries in Italy depend only on academic rank and seniority.

As for the cost of capital, *k*, we rely on data at discipline level taken from Abramo, Aksnes, and D'Angelo (2020). Table 2 summarizes cost of capital (k), total cost of production factors ($\frac{w_r}{2}+\text{k}$), and total cost normalization factors per academic rank and discipline. The normalization factor used is the minimum recorded value of total cost, occurring for assistant professors in Psychology (54,081 Euro).

In what follows, we use the total cost normalization factors to report measures of productivity.

***Table 2: Average yearly cost of production factors, by academic rank and discipline, for Italian professors***

| | | $\frac{w_r}{2}+\text{k}$ | | | Total cost normalization factor | | |
|---|---|---|---|---|---|---|---|
| UDA | k | Assistant professors | Associate professors | Full professors | Assistant professors | Associate professors | Full professors |
| 1 - Mathematics and computer science | 30822 | 58109 | 65079 | 82019 | 1.07 | 1.20 | 1.52 |
| 2 - Physics | 46194 | 73481 | 80451 | 97391 | 1.36 | 1.49 | 1.80 |
| 3 - Chemistry | 39820 | 67107 | 74077 | 91017 | 1.24 | 1.37 | 1.68 |
| 4 - Earth sciences | 60016 | 87303 | 94273 | 111213 | 1.61 | 1.74 | 2.06 |
| 5 - Biology | 45748 | 73035 | 80005 | 96945 | 1.35 | 1.48 | 1.79 |
| 6 - Medicine | 41228 | 68515 | 75485 | 92424 | 1.27 | 1.40 | 1.71 |
| 7 - Agricultural and veterinary sciences | 45748 | 73035 | 80005 | 96945 | 1.35 | 1.48 | 1.79 |
| 8 - Civil engineering | 47810 | 75097 | 82067 | 99007 | 1.39 | 1.52 | 1.83 |
| 9 - Industrial and information engineering | 47810 | 75097 | 82067 | 99007 | 1.39 | 1.52 | 1.83 |
| 10 - Psychology | 26777 | 54064 | 61034 | 77974 | 1.00 | 1.13 | 1.44 |

We assume that capital is the same, in both periods, across individuals and institutions within UDAs, as we lack specific information. It may be objected that high performing universities received more funds in the second period, however we do not know what share of these funds were allocated to other uses or to research, or if regional governments compensated for the lower national funding to their own low-performing universities.

Given the different intensity of publication across research fields, the comparison of performance for universities operating in heterogeneous fields cannot be achieved by simply averaging individual performance across fields. A two-step procedure is required: first measuring the productivity of the individual professors in their field (formula 1), and then normalizing individual productivity by the average in the field. At the aggregate level of university, UDA, and SDS, the yearly productivity *FSS* for the aggregate unit is:

both first and last author; 15% to both second and last but one author; and the remaining 10% is divided among all others. The weighting values were assigned following advice from senior Italian professors in the life sciences. The values could be changed to suit different practices in other national contexts.

[13] https://dalia.cineca.it/php4/inizio_access_cnvsu.php, last accessed on 24 November, 2020.

$$FSS = \frac{1}{RS}\sum_{j=1}^{RS}\frac{FSS_{Pj}}{\overline{FSS_P}}$$
[2]

Where:
$RS$ = number of professors in the unit, in the observed period;
$FSS_{Pj}$= productivity of professor $j$ in the unit;
$\overline{FSS_P}$= average productivity of all Italian productive professors under observation in the same SDS of professor $j$.

The $FSS$ at aggregate level is therefore a relative performance indicator. For instance, an $FSS$ value of 1.10 means that the unit's performance is 10% above average.

Because of lack of data, the proposed analysis does not account for possible changes in research productivity confounding factors, such as amount of resources available for research, working conditions, etc. (Dundar & Lewis, 1998).

## 5. Results

We first examine the results of the ranking comparison at the overall university level in the two periods. Since rank variations are likely to be different for each university across UDAs and SDSs within each single UDA, we then also examine rank differences at UDA and SDS levels: analyses that could serve management in formulating their investment, recruitment and competitive strategies up to the levels of individual SDSs.

Figure 1 shows the overall relative research performance of the 60 universities with at least 30 professors assessable in both five-year periods, in decreasing order of change in ranking.[14]

Table 3 shows the analytical data of performance scores and ranks in the two periods, and notes the macro-region locality of the universities. The last column of the table indicates the relative rank variation, i.e. with respect to maximum possible variation for the individual university. The relative Δrank is therefore the ratio between the obtained rank shift and the maximum obtainable rank shift.

For University Vita-Salute San Raffaele of Milan, at the top in both periods, a positive Δrank would have been impossible. For University of Salerno, the 14 positions gained represent 74% of the 19 scalable positions between the two periods. Following this, in order, place the four universities of Naples 'Federico II', of Salento, the 'Parthenope' University, and the SISSA of Trieste. The latter gained only one position, but of only two possible, considering its third place in the first period. The SISSA is also the only northern university among the first ten on the list: of the remaining nine, two are in central Italy and seven in the south. The universities with the worst relative negative rank jumps are Udine, Siena and the University of Basilicata.

[14] Differently form the VQR, we do not classify and rank universities into categories by size. The reason is that neither differing returns to size (Abramo, Cicero, & D'Angelo, 2012; Bonaccorsi & Daraio, 2005; Seglen & Aksnes, 2000; Golden & Carstensen, 1992), nor returns to scope seem to occur in academic research (Abramo, D'Angelo, & Di Costa, 2014b).

*Figure 1: For the 60 universities with at least 30 professors in the in the fields under observation: variation in performance (FSS) rank, 2013-2017 vs 2007-2011*

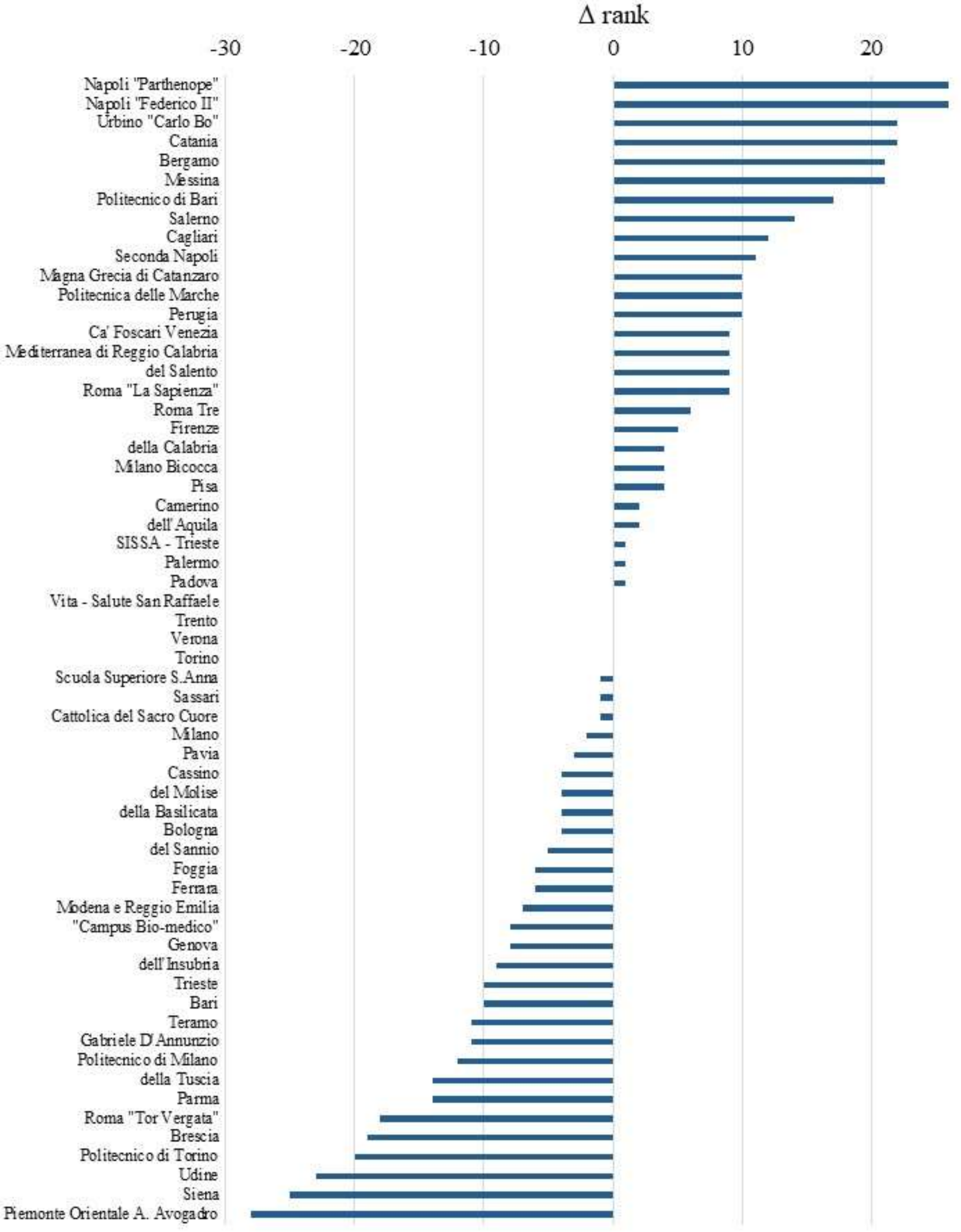

*Table 3: For the 60 universities observed: productivity (FSS) scores, ranks and relative Δrank, 2013-2017 vs 2007-2011*

| | | 2007-2011 | | | 2013-2017 | | | | | |
|---|---|---|---|---|---|---|---|---|---|---|
| University | Macro region† | Staff | FSS | rank | Staff | FSS | rank | ΔFSS | Δrank | Relative Δrank |
| Vita - Salute San Raffaele | N | 63 | 3.27 | 1 | 85 | 2.42 | 1 | -0.85 | 0 | n.a. |
| Salerno | S | 448 | 0.97 | 20 | 417 | 1.22 | 6 | 0.25 | 14 | 74% |
| Napoli ‘Federico II’ | S | 1580 | 0.77 | 42 | 1453 | 1.04 | 16 | 0.27 | 26 | 63% |
| del Salento | S | 261 | 1.00 | 18 | 228 | 1.11 | 9 | 0.11 | 9 | 53% |
| Napoli ‘Parthenope’ | S | 122 | 0.65 | 51 | 118 | 1.00 | 25 | 0.35 | 26 | 52% |
| SISSA - Trieste | N | 50 | 1.72 | 3 | 60 | 1.85 | 2 | 0.13 | 1 | 50% |
| Urbino ‘Carlo Bo’ | C | 148 | 0.65 | 50 | 121 | 0.96 | 28 | 0.31 | 22 | 45% |
| Catania | S | 823 | 0.63 | 52 | 672 | 0.95 | 30 | 0.32 | 22 | 43% |
| Politecnico di Bari | S | 219 | 0.77 | 43 | 179 | 0.96 | 26 | 0.20 | 17 | 40% |
| Messina | S | 717 | 0.62 | 53 | 618 | 0.92 | 32 | 0.30 | 21 | 40% |
| Politecnica delle Marche | C | 385 | 0.93 | 27 | 365 | 1.03 | 17 | 0.10 | 10 | 38% |
| Milano Bicocca | N | 458 | 1.11 | 12 | 436 | 1.14 | 8 | 0.03 | 4 | 36% |
| Bergamo | N | 90 | 0.48 | 60 | 86 | 0.86 | 39 | 0.38 | 21 | 36% |
| Firenze | C | 1022 | 1.02 | 17 | 878 | 1.09 | 12 | 0.07 | 5 | 31% |
| Perugia | C | 679 | 0.85 | 34 | 596 | 1.00 | 24 | 0.15 | 10 | 30% |
| Roma Tre | C | 252 | 0.94 | 24 | 245 | 1.03 | 18 | 0.09 | 6 | 26% |
| Seconda Napoli | S | 585 | 0.73 | 44 | 524 | 0.92 | 33 | 0.19 | 11 | 26% |
| Magna Grecia di Catanzaro | S | 157 | 0.78 | 41 | 138 | 0.93 | 31 | 0.15 | 10 | 25% |
| Cagliari | S | 573 | 0.61 | 55 | 498 | 0.82 | 43 | 0.21 | 12 | 22% |
| della Calabria | S | 439 | 0.98 | 19 | 409 | 1.04 | 15 | 0.07 | 4 | 22% |
| Roma ‘La Sapienza’ | C | 2339 | 0.69 | 45 | 1945 | 0.87 | 36 | 0.18 | 9 | 20% |
| Padova | N | 1335 | 1.36 | 6 | 1271 | 1.26 | 5 | -0.10 | 1 | 20% |
| Ca’ Foscari Venezia | N | 103 | 0.68 | 47 | 85 | 0.86 | 38 | 0.18 | 9 | 20% |
| Mediterranea di Reggio Calabria | S | 139 | 0.65 | 49 | 137 | 0.85 | 40 | 0.20 | 9 | 19% |
| Pisa | C | 955 | 0.95 | 23 | 841 | 1.02 | 19 | 0.07 | 4 | 18% |
| Camerino | C | 189 | 0.60 | 57 | 192 | 0.75 | 55 | 0.15 | 2 | 4% |
| dell’Aquila | S | 425 | 0.52 | 59 | 362 | 0.75 | 57 | 0.23 | 2 | 3% |
| Palermo | S | 963 | 0.68 | 46 | 848 | 0.82 | 45 | 0.13 | 1 | 2% |
| Trento | N | 239 | 1.51 | 4 | 238 | 1.40 | 4 | -0.11 | 0 | 0% |
| Verona | N | 349 | 1.33 | 7 | 326 | 1.16 | 7 | -0.17 | 0 | 0% |
| Torino | N | 1078 | 1.14 | 11 | 1010 | 1.10 | 11 | -0.05 | 0 | 0% |
| Scuola Superiore S.Anna | C | 43 | 2.16 | 2 | 53 | 1.80 | 3 | -0.36 | -1 | -2% |
| Cattolica del Sacro Cuore | N | 740 | 0.96 | 22 | 618 | 1.00 | 23 | 0.04 | -1 | -3% |
| Milano | N | 1358 | 1.22 | 8 | 1210 | 1.10 | 10 | -0.11 | -2 | -4% |
| Bologna | N | 1542 | 1.14 | 10 | 1385 | 1.06 | 14 | -0.08 | -4 | -8% |
| Pavia | N | 604 | 0.93 | 26 | 512 | 0.95 | 29 | 0.03 | -3 | -9% |
| Ferrara | N | 389 | 1.07 | 14 | 349 | 1.02 | 20 | -0.05 | -6 | -13% |
| Foggia | S | 158 | 1.06 | 16 | 146 | 1.00 | 22 | -0.06 | -6 | -14% |
| ‘Campus Bio-medico’ | C | 78 | 1.38 | 5 | 96 | 1.09 | 13 | -0.30 | -8 | -15% |
| del Sannio | S | 108 | 0.90 | 29 | 108 | 0.91 | 34 | 0.01 | -5 | -16% |
| Politecnico di Milano | N | 819 | 1.18 | 9 | 835 | 1.01 | 21 | -0.17 | -12 | -24% |
| Modena e Reggio Emilia | N | 523 | 0.84 | 35 | 470 | 0.83 | 42 | -0.01 | -7 | -28% |
| dell’Insubria | N | 248 | 0.92 | 28 | 218 | 0.86 | 37 | -0.06 | -9 | -28% |
| Genova | N | 746 | 0.86 | 33 | 699 | 0.85 | 41 | -0.01 | -8 | -30% |
| della Tuscia | C | 145 | 1.09 | 13 | 151 | 0.96 | 27 | -0.13 | -14 | -30% |
| Cassino | C | 116 | 0.68 | 48 | 104 | 0.77 | 52 | 0.09 | -4 | -33% |
| Trieste | N | 392 | 0.84 | 37 | 343 | 0.81 | 47 | -0.03 | -10 | -43% |
| Politecnico di Torino | N | 583 | 1.06 | 15 | 576 | 0.90 | 35 | -0.16 | -20 | -44% |
| Bari | S | 848 | 0.82 | 38 | 748 | 0.81 | 48 | -0.01 | -10 | -45% |
| Sassari | S | 375 | 0.54 | 58 | 326 | 0.67 | 59 | 0.13 | -1 | -50% |
| Parma | N | 625 | 0.86 | 32 | 533 | 0.81 | 46 | -0.05 | -14 | -50% |
| Teramo | S | 89 | 0.79 | 39 | 89 | 0.79 | 50 | 0.00 | -11 | -52% |
| Brescia | N | 380 | 0.94 | 25 | 359 | 0.82 | 44 | -0.12 | -19 | -54% |
| Gabriele D’Annunzio | S | 331 | 0.78 | 40 | 307 | 0.78 | 51 | 0.00 | -11 | -55% |
| del Molise | S | 129 | 0.62 | 54 | 117 | 0.68 | 58 | 0.06 | -4 | -67% |
| Piemonte Orientale A. Avogadro | N | 186 | 0.97 | 21 | 174 | 0.81 | 49 | -0.17 | -28 | -72% |
| Roma ‘Tor Vergata’ | C | 908 | 0.84 | 36 | 775 | 0.76 | 54 | -0.08 | -18 | -75% |
| Udine | N | 370 | 0.89 | 30 | 358 | 0.76 | 53 | -0.13 | -23 | -77% |
| Siena | C | 451 | 0.88 | 31 | 330 | 0.75 | 56 | -0.14 | -25 | -86% |
| della Basilicata | S | 225 | 0.60 | 56 | 219 | 0.64 | 60 | 0.03 | -4 | -100% |

† *N, north; C, center; S, south*

For a more immediate understanding of the relative performance variations of universities from period to period, Figure 2 shows the performance plots (columns 3 and 6, Table 3). In the second plot, the tails have clearly drawn nearer, indicating a reduction of performance dispersion: the 20 bottom ranked universities for *FSS* 2007-2011 have all increased their relative performance, however among the top 11 ranked universities, only SISSA of Trieste improves, from *FSS* 1.724 to 1.854.

In general terms, the score dispersion between the two periods appears significantly reduced (their standard deviations are, respectively, 0.425 and 0.290). To verify whether the convergence that we observe is statistically significant, we apply the Fisher variance comparison test to the variances of the before- and after-distribution, obtaining an *F* value equal to 2.140, which is significant with *p* values below 0.5%.

The convergence is in line with that reported by Checchi, Mazzotta, Momigliano, & Olivanti (2020), who analyze the variation in performance scores according to VQR criteria.

As noted in the introduction, the negative Δ*FSS* of top performing universities should not be immediately interpreted as a reduction in their faculty's absolute productivity. In fact, although this may have occurred, the reductions are also certainly caused by the increase in average productivity of all national professors. Those on the left tail have contributed most to increasing the scaling value of sectoral distributions, and consequently, induce a reduction in the normalized *FSS* for the professors on the right.

***Figure 2: For the 60 universities observed: productivity (FSS) values, 2013-2017 vs 2007-2011***

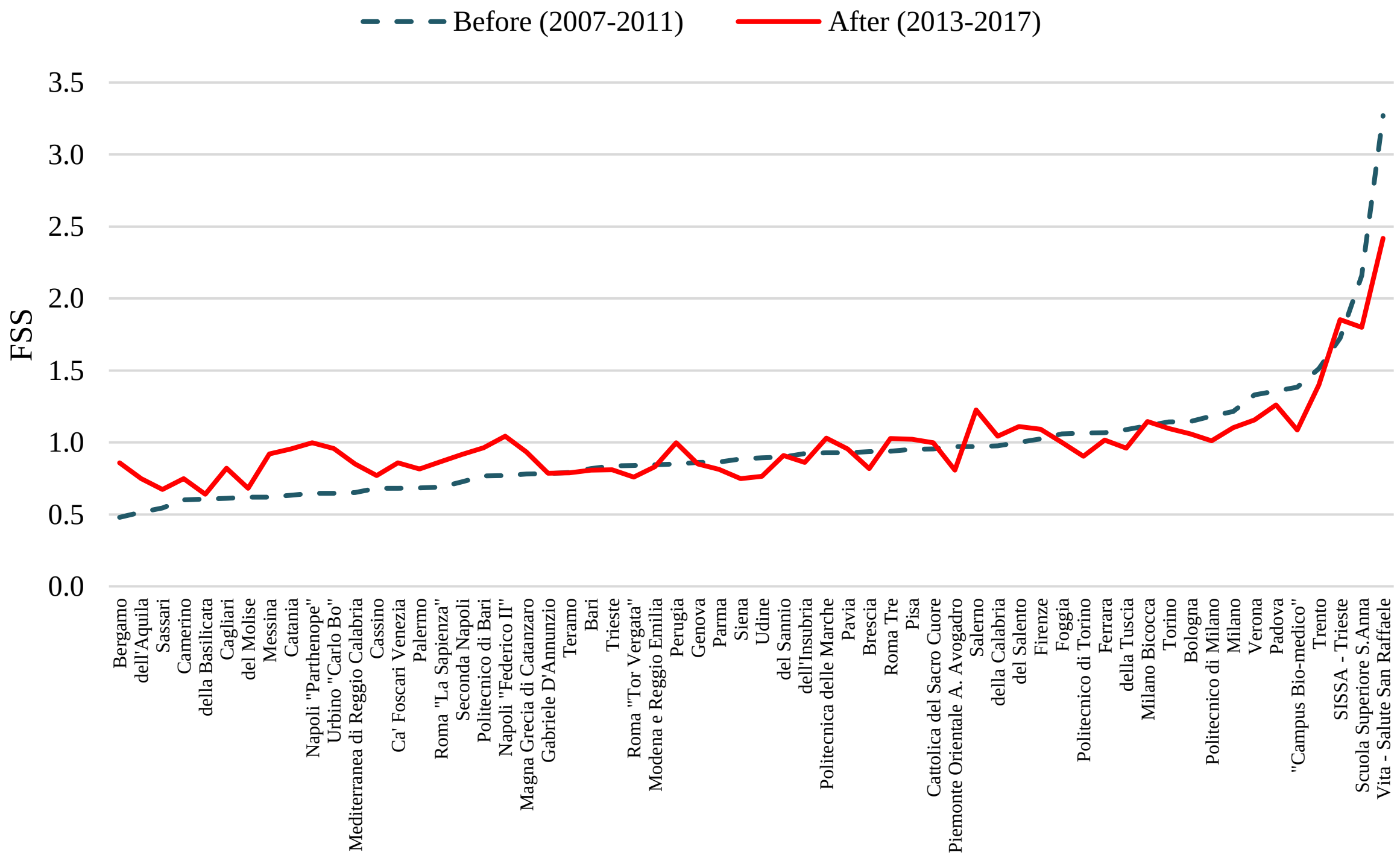

The distribution of *FSS* for universities shows greater variability in the first period than the second, as shown by range of variation (max-min: 2.791 vs 1.777), interquartile deviation (Q3-Q1: 0.372 vs 0.228), standard deviation (0.425 vs 0.290), and skewness (3.261 vs 2.874). The average and median *FSS* of the second period are slightly higher than those of the first period. Because at aggregate level the *FSS* is the average of the "scaled" values (where the scaling factor is the average of SDS distributions, zeros removed), it should be noted that this means that the difference in averages/mediums between the two periods can only be due to the lower incidence of unproductive professors, i.e. zero *FSS* values, which is actually 8.3% in 2007-2011 and 2.9% in 2013-2017.

Table 4 provides a summary of the analysis of the relative change in performance at the overall level of the universities, broken down by macro-region. Of the 24 universities in the north, only 5 (21%) improved their position in the national rank, compared to 64% of the universities in the south (14 out of 22). Also in relative terms, the performance of the southern universities seems to have improved significantly, considering that the average value of their relative Δrank is positive and equal to +4%, compared to -14% of the northern universities and -2% of those in the center.

***Table 4: Change in performance (FSS) rank over the two periods, for universities in each macro region***

| Macro region | No. of universities | Of which - improve | Of which - worsen | Avg . rel. Δrank | Max improvement | Max. decline |
|---|---|---|---|---|---|---|
| North | 24 | 5 | 15 | -14% | 50% | -77% |
| Center | 14 | 8 | 6 | -2% | 45% | -86% |
| South | 22 | 14 | 8 | 4% | 74% | -100% |

To better grasp to what extent the relative productivity variations depend on the first period productivity and on the localization of the universities, we run an OLS model (Table 5), where the first variable is the first period productivity and the second is a dummy, whose value is 1 if the university is located in the south and 0 otherwise.

The general convergence of performance is shown by the sign and significance of the first regressor (-0.344), which indicates that the productivity increase is inversely correlated to the relative productivity in the first period. Alongside, the coefficient of the dummy "south" shows a statistically significant evidence that on average, universities in the south, increased their FSS score remarkably more than universities from the other two macro regions. This higher increase seems to be additional to the average convergence effect.

***Table 5: Estimation results for linear regression model (dependent variable ΔFSS)***

| | Coeff. | Std Err. | t | P>t | [95% Conf. interval] | |
|---|---|---|---|---|---|---|
| Const. | 0.333 | 0.044 | 7.630 | 0.000 | 0.245 | 0.420 |
| $FSS_{2007\text{-}2011}$ | -0.344 | 0.040 | -8.710 | 0.000 | -0.423 | -0.265 |
| South | 0.069 | 0.031 | 2.200 | 0.032 | 0.006 | 0.131 |

*Number of obs = 60; F(2,57)=51.50; Prob>F=0.000*
*R-squared=0.675, RootMSE=0.114*

For each university we also analyze the relative performance variation in each UDA. Table 6 presents the case of University of Firenze (Florence). Between the two periods examined, out of 10 active UDAs, this university improved its position in four (Physics; Biology; Industrial and information engineering; Psychology), worsened in three (Mathematics and computer science; Agricultural and veterinary sciences; Civil engineering) and maintained unchanged rank in Chemistry (maintaining the national top spot), Earth sciences (second place) and Medicine (eleventh).

***Table 6: Productivity (FSS) scores and ranks of the University of Florence in the two periods considered, by UDA***

| | 2007-2011 | | | 2013-2017 | | | | |
|---|---|---|---|---|---|---|---|---|
| UDA | Staff | FSS | rank | Staff | FSS | rank | Δ rank | Δ FSS |
| 1 - Mathematics and computer science | 104 | 0.87 | 24 | 76 | 0.77 | 35 | -11 | -0.10 |
| 2 - Physics | 71 | 0.68 | 36 | 58 | 0.72 | 33 | +3 | +0.03 |
| 3 - Chemistry | 109 | 2.02 | 1 | 99 | 1.87 | 1 | 0 | -0.16 |
| 4 - Earth sciences | 37 | 1.56 | 2 | 37 | 1.72 | 2 | 0 | +0.17 |
| 5 - Biology | 124 | 1.20 | 10 | 121 | 1.15 | 8 | +2 | -0.05 |
| 6 - Medicine | 272 | 1.13 | 11 | 220 | 1.09 | 11 | 0 | -0.05 |
| 7 - Agricultural and veterinary sciences | 119 | 0.62 | 20 | 96 | 0.71 | 21 | -1 | +0.09 |
| 8 - Civil engineering | 45 | 0.45 | 26 | 43 | 0.58 | 30 | -4 | +0.13 |
| 9 - Industrial and information engineering | 99 | 0.50 | 44 | 92 | 0.95 | 22 | +22 | +0.44 |
| 10 - Psychology | 42 | 0.67 | 11 | 36 | 1.40 | 1 | +10 | +0.74 |

Table 7 shows, for each of the 60 universities, the number of UDAs evaluated, the percentage of UDAs with positive variation in rank between the two periods, as well as the specific UDAs that improved, worsened or remained unchanged in national positioning.

Among the very large universities, evaluable in all 10 UDAs considered, Naples 'Federico II' places at the top, having improved in all disciplines but Agricultural and veterinary sciences. Rome 'La Sapienza' also does very well, improving in 8 UDAs. Padua and Bologna contrast, dropping in national rank in 7 out of 10 UDAs. Catania, a southern university with 9 active UDAs, also improves in all of them. On the opposite front are Trento and Politecnico di Milano, which do not improve in any of their 5 active UDAs, and instead worsen in four.

***Table 7: Variation in performance (FSS) rank of the universities in the two periods, per UDA***

| | | | | UDA* | | | | | | | | | |
|---|---|---|---|---|---|---|---|---|---|---|---|---|---|
| University | Macro region† | Assessed UDAs | Of which increasing | 1 | 2 | 3 | 4 | 5 | 6 | 7 | 8 | 9 | 10 |
| Napoli 'Federico II' | S | 10 | 90% | + | + | + | + | + | + | - | + | + | + |
| Roma 'La Sapienza' | C | 10 | 80% | + | + | + | - | + | + | - | + | + | + |
| Messina | S | 10 | 60% | - | - | + | = | + | + | + | - | + | + |
| Palermo | S | 10 | 60% | + | - | + | + | + | - | + | - | + | - |
| Firenze | C | 10 | 40% | - | + | = | = | + | = | - | - | + | + |
| Parma | N | 10 | 30% | - | - | - | - | - | + | + | - | - | + |
| Bologna | N | 10 | 20% | - | + | - | - | + | = | - | - | - | - |
| Padova | N | 10 | 20% | - | + | - | = | + | - | - | - | - | - |
| Catania | S | 9 | 100% | + | + | + | + | + | + | + | + | + | |
| Pavia | N | 9 | 78% | + | + | + | + | - | - | | + | + | + |
| Cagliari | S | 9 | 67% | + | + | = | + | - | + | | + | + | - |
| Perugia | C | 9 | 56% | + | - | + | - | - | + | + | - | + | |
| Genova | N | 9 | 44% | + | + | - | + | - | = | | - | - | + |
| Modena e Reggio Emilia | N | 9 | 44% | - | - | + | + | - | + | - | | + | - |
| Milano | N | 9 | 33% | - | - | + | - | - | + | = | | + | - |
| Pisa | C | 9 | 33% | - | + | - | - | + | = | + | - | - | |
| Trieste | N | 9 | 22% | - | - | - | + | - | - | | + | - | - |
| Bari | S | 9 | 11% | - | - | + | - | - | - | = | | - | - |
| della Calabria | S | 8 | 50% | - | + | - | + | + | - | | + | - | |
| Ferrara | N | 8 | 50% | + | + | + | + | - | - | | - | - | |
| Seconda Napoli | S | 8 | 50% | - | - | + | | + | + | | + | - | - |
| Milano Bicocca | N | 8 | 38% | + | + | - | = | - | + | | | - | - |
| Politecnica delle Marche | C | 8 | 38% | - | - | = | | + | - | + | - | + | |
| Torino | N | 8 | 38% | + | = | - | + | - | + | - | | | - |
| della Basilicata | S | 8 | 25% | - | + | = | - | + | | - | = | - | |
| Udine | N | 8 | 25% | - | - | + | | + | - | - | - | - | |
| Salerno | S | 7 | 86% | + | + | - | | + | + | | + | + | |
| dell’Aquila | S | 7 | 71% | - | + | - | | + | + | | + | + | |
| Camerino | C | 7 | 57% | = | - | + | + | + | + | - | | | |
| Gabriele D’Annunzio | S | 7 | 29% | - | | + | - | = | - | | = | | + |
| Roma ‘Tor Vergata’ | C | 7 | 14% | - | + | - | | - | - | | - | - | |
| Siena | C | 7 | 14% | = | = | - | - | - | - | | | + | |
| Cattolica del Sacro Cuore | N | 6 | 67% | + | + | | | - | - | + | | | + |
| Roma Tre | C | 6 | 67% | - | + | | = | + | | | + | + | |
| Verona | N | 6 | 33% | - | | | | - | - | + | | + | - |
| Brescia | N | 6 | 17% | + | - | | | - | - | | - | - | |
| del Salento | S | 5 | 60% | - | - | + | | + | | | | + | |
| Piemonte Orientale A. Avogadro | N | 5 | 40% | + | + | - | | - | - | | | | |
| dell’Insubria | N | 5 | 20% | - | - | - | | - | + | | | | |
| Politecnico di Torino | N | 5 | 20% | - | - | + | | | | | - | - | |
| Politecnico di Milano | N | 5 | 0% | = | - | - | | | | | - | - | |
| Trento | N | 5 | 0% | - | - | | | | | | - | = | - |
| Napoli ‘Parthenope’ | S | 4 | 75% | + | | | | = | | | + | + | |
| Politecnico di Bari | S | 4 | 75% | + | + | | | | | | - | + | |
| Sassari | S | 4 | 75% | | | + | | + | - | + | | | |
| Mediterranea di Reggio Calabria | S | 4 | 50% | - | | | | | | + | - | + | |
| Urbino ‘Carlo Bo’ | C | 4 | 25% | | | - | + | - | | | | | - |
| Bergamo | N | 3 | 67% | - | | | | | | | | + | + |
| del Sannio | S | 3 | 67% | | | | - | + | | | | + | |
| Foggia | S | 3 | 67% | | | | | + | - | + | | | |
| Magna Grecia di Catanzaro | S | 3 | 67% | | | - | | + | + | | | | |
| del Molise | S | 3 | 33% | | | | | - | + | = | | | |
| della Tuscia | C | 3 | 0% | | | - | | - | | - | | | |
| Ca’ Foscari Venezia | N | 2 | 100% | + | | + | | | | | | | |
| ‘Campus Bio-medico’ | C | 2 | 50% | | | | | | - | | | + | |
| Cassino | C | 2 | 50% | | | | | | | | + | - | |
| SISSA - Trieste | N | 2 | 50% | - | + | | | | | | | | |
| Vita - Salute San Raffaele | N | 2 | 0% | | | | | = | = | | | | |
| Scuola Superiore S.Anna | C | 1 | 0% | | | | | | | | | = | |
| Teramo | S | 1 | 0% | | | | | | | - | | | |

† *N, north; C, center; S, south*

* *1 - Mathematics and computer science, 2 - Physics, 3 - Chemistry, 4 - Earth sciences, 5 - Biology, 6 - Medicine, 7 - Agricultural and veterinary sciences, 8 - Civil engineering, 9 - Industrial and information engineering, 10 - Psychology*

We repeat the analysis for each of the 199 SDSs considered. Table 8 presents, by way of example, the analysis of the Physics SDSs for the University of Florence, while Table 9 summarizes the results of this type of analysis for all universities active in this UDA.

***Table 8: Productivity (FSS) scores and ranks of University of Firenze, in the two periods considered, by SDSs of Physics***

| | 2007-2011 | | | 2013-2017 | | | | |
|---|---|---|---|---|---|---|---|---|
| SDS | Staff | FSS | rank | Staff | FSS | rank | Δ rank | Δ FSS |
| FIS/01 | 19 | 0.62 | 35 | 11 | 0.72 | 30 | +5 | +0.09 |
| FIS/02 | 10 | 0.34 | 33 | 9 | 0.52 | 31 | +2 | +0.18 |
| FIS/03 | 20 | 0.94 | 14 | 23 | 0.74 | 24 | -10 | -0.21 |
| FIS/04 | 4 | 0.94 | 11 | 3 | 1.62 | 6 | +5 | +0.68 |
| FIS/05 | 9 | 0.60 | 17 | 5 | 0.59 | 18 | -1 | -0.01 |
| FIS/07 | 9 | 0.46 | 33 | 6 | 0.31 | 38 | -5 | -0.15 |

*FIS/01, Experimental Physics; FIS/02, Theoretical Physics, Mathematical Models and Methods; FIS/03, Physics of matter; FIS/04, Nuclear and Subnuclear Physics; FIS/05, Astronomy and Astrophysics; FIS/06, Physics for Earth and Atmospheric Sciences; FIS/07, Applied Physics (Cultural Heritage, Environment, etc.); FIS/08, Didactics and History of Physics*

***Table 9: Variation in the performance (FSS) rank of the universities in the two periods, for the SDSs of the Physics UDA***

| University | Macro region† | Assessed SDSs | Of which increasing | FIS/01 | FIS/02 | FIS/03 | FIS/04 | FIS/05 | FIS/06 | FIS/07 | FIS/08 |
|---|---|---|---|---|---|---|---|---|---|---|---|
| Bologna | N | 8 | 50% | + | + | + | - | - | = | - | + |
| Milano | N | 8 | 50% | + | = | - | - | + | - | + | + |
| Napoli ‘Federico II’ | S | 8 | 50% | + | + | + | - | - | = | + | - |
| Ferrara | N | 7 | 71% | + | + | - | + | - | + | + | |
| Milano Bicocca | N | 7 | 71% | + | - | + | + | = | + | + | |
| Roma ‘La Sapienza’ | C | 7 | 71% | + | + | - | + | + | - | + | |
| Torino | N | 7 | 57% | + | - | + | + | - | - | + | |
| Roma Tre | C | 7 | 43% | + | - | + | + | = | = | - | |
| Padova | N | 7 | 29% | - | + | + | - | - | | - | - |
| del Salento | S | 7 | 14% | - | + | - | - | - | - | - | |
| Catania | S | 6 | 67% | + | + | - | - | + | | + | |
| dell’Aquila | S | 6 | 67% | + | + | + | + | | - | - | |
| Pavia | N | 6 | 67% | + | - | + | + | | | + | - |
| Roma ‘Tor Vergata’ | C | 6 | 67% | + | - | + | - | + | | + | |
| Cagliari | S | 6 | 50% | + | - | - | + | - | | + | |
| Firenze | C | 6 | 50% | + | + | - | + | - | | - | |
| Trento | N | 6 | 50% | - | - | + | - | | + | | + |
| Trieste | N | 6 | 50% | - | = | - | + | + | | + | |
| della Calabria | S | 6 | 33% | - | + | + | | - | = | - | |
| Modena e Reggio Emilia | N | 6 | 33% | - | + | - | + | | - | = | |
| Palermo | S | 6 | 33% | - | + | + | | - | | - | - |
| Pisa | C | 6 | 33% | + | - | - | - | = | | + | |
| dell’Insubria | N | 6 | 17% | + | - | - | - | = | | = | |
| Perugia | C | 6 | 17% | + | - | - | - | - | | - | |
| Bari | S | 5 | 60% | + | - | | | | = | + | + |
| Genova | N | 5 | 60% | + | + | - | + | | | - | |
| Udine | N | 5 | 20% | - | - | - | | | | - | + |
| Cattolica del Sacro Cuore | N | 4 | 75% | + | - | + | | | | + | |
| Piemonte Orientale A. Avogadro | N | 4 | 75% | + | - | | | | + | + | |
| Salerno | S | 4 | 75% | + | - | + | | | | + | |
| Scuola Normale Superiore | C | 4 | 75% | + | + | = | | + | | | |
| Parma | N | 4 | 50% | - | + | + | | | | - | |
| Politecnico di Torino | N | 4 | 25% | - | - | + | - | | | | |
| Seconda Napoli | S | 4 | 0% | - | - | - | | | | - | |
| SISSA - Trieste | N | 3 | 100% | | + | + | | + | | | |
| Camerino | C | 3 | 67% | + | + | - | | | | | |
| Messina | S | 3 | 67% | + | | - | | | | + | |
| Brescia | N | 3 | 33% | - | | - | | | | + | |
| della Basilicata | S | 3 | 33% | + | = | | | | = | | |
| Urbino ‘Carlo Bo’ | C | 3 | 33% | + | | | | | - | | = |
| Napoli ‘Parthenope’ | S | 2 | 100% | + | | | | + | | | |
| Politecnico di Bari | S | 2 | 100% | + | | + | | | | | |
| Siena | C | 2 | 50% | - | | | | | | + | |
| Verona | N | 2 | 50% | - | | | | | | + | |
| Ca’ Foscari Venezia | N | 2 | 0% | - | | - | | | | | |
| Politecnica delle Marche | C | 2 | 0% | - | | | | | | - | |
| Politecnico di Milano | N | 2 | 0% | - | | - | | | | | |
| ‘Campus Bio-medico’ | C | 1 | 100% | | + | | | | | | |
| del Sannio | S | 1 | 100% | + | | | | | | | |
| Guglielmo Marconi (on line) | C | 1 | 100% | + | | | | | | | |
| IUSS | N | 1 | 100% | | | | | + | | | |
| Sassari | S | 1 | 100% | | | | | | | + | |
| Cassino | C | 1 | 0% | - | | | | | | | |
| della Tuscia | C | 1 | 0% | | | | | | | - | |
| Enna - KORE | S | 1 | 0% | - | | | | | | | |
| Foggia | S | 1 | 0% | | | | | | | - | |
| Gabriele D’Annunzio | S | 1 | 0% | | | | | | | - | |
| Mediterranea di Reggio Calabria | S | 1 | 0% | - | | | | | | | |
| Teramo | S | 1 | 0% | | | | | | | - | |
| UniNettuno (on line) | C | 1 | 0% | - | | | | | | | |

*† N, north; C, center; S, south*

*FIS/01, Experimental Physics; FIS/02, Theoretical Physics, Mathematical Models and Methods; FIS/03, Physics of matter; FIS/04, Nuclear and Subnuclear Physics; FIS/05, Astronomy and Astrophysics; FIS/06, Physics for Earth and Atmospheric Sciences; FIS/07, Applied Physics (Cultural Heritage, Environment, etc.); FIS/08, Didactics and History of Physics*

## 6. Conclusions

Individuals and organizations respond differently to financial and reputational incentives. In the case of national PBRFS, the response depends on the extent that the organizations internally deploy the incentives at individual level, but also on the researchers’ sense of belonging and identification with the cause or “distinguishing vision” of their organization. In this paper we have witnessed the truth of these premises, thanks to the assessment of the variation in the relative performance of Italian universities, straddling the first ever two editions of a national research assessment exercise.

In different national contexts, PBRFS is generally adopted with the main aim of improving scientific productivity. The current study, dealing only with relative performance, does not ascertain whether the Italian VQR indeed reaches that objective. However, in the case of either across-the-board improvement, or across-the-board worsening, we can certainly say that some universities have improved much more than others, or that some universities have worsened much less than others. The truth is probably that some have improved their absolute performance while others have worsened: this verification, which is much more complex, will be addressed in a future study.

We find that before introduction of PBRFS the universities of the south were much less productive than those of the north. Then, sparked by the financial incentive of PBRFS or perhaps by questions of pride, they reacted to the VQR more strongly than the others, gaining positions remarkably.

This result is remarkable, in fact southern universities, because of their low performance in the VQR, received on average relatively lower government funds than others. Nor were there other policy interventions that would support such relative increase in research productivity. We conclude then that the VQR-based PBRFS had no displacement effect on low performing universities, in terms of research performance. At the same time though, the effect on reputation revealed in the increased migration of southern students to universities in the north, as shown by enrollment statistics (ANVUR, 2014, p. 37). From a different perspective, this also reveals how poor the productivity of the southern universities was before the VQR.

The result is certainly positive in terms of reducing the north-south divide in infrastructure and economic activity, long present in Italy. A particular benefit is that the relative improvement of southern universities could slow the phenomenon of northward student migration, in search of better education, and mitigate problems of social discrimination at regional and national levels. In fact it is only the richer families who can support the costs of leaving for university studies, especially to the North, where living costs are much higher.

Our study also observes the convergence of performance by the universities: between 2007-2011 and 2013-2017, the distances diminished between Italy's most and least productive. If the VQR was designed as a stimulus to the development of Italian "champion universities", through selective funding, then a reconsideration could be in order. In any case, our subsequent study will check for variations in absolute performance of the individual universities.

Evaluative bibliometric analysis, such as provided here, also serves at the level of the individual research organizations, to inform strategic management. In this case, each university can understand where it stands, and over time know how much it has improved or worsened compared to its "competitors", in each discipline and field. The individual universities can assess which of any internal policies and actions have worked, and further refine their competitive strategies based on probing examination.

Finally, in our opinion, this work represents a significant advance in the literature. First of all, for the performance indicator used - which does not separate the publication from its impact, accounts for the contributions of individual authors, and as far as possible, the resources used for production - but also for the very broad observation of national production, unlike the small fraction used in many other examinations of PBRFS.

Last but not least, because of the fine-grained classification scheme of the professors, we are able to control for the different intensity of publication between fields, and greatly limit the distortions occurring in direct comparison of performance between researchers in different fields. This classification scheme, to our knowledge unique in the world, allows the benefit of precise and penetrating diagnoses, and therefore unique possibilities of informing micro-level policies and strategies.

As with other scientometric investigations, there remain some limits concerning availability of data, first of all on the universities' allocation of resources to the individual researchers. Second, the new knowledge produced is not always embedded in publications, and the bibliographic repertories (such as WoS, used here) do not register all publications. Third, the measurement of the impact of publications using citation-based indicators is a prediction, not definitive. Fourth, citations can also be negative or inappropriate, and in any case they certify only scholarly impact, forgoing other types of impact. Fifth, given the limitations on input data (e.g. costs of labor and capital), we make several assumptions in measuring professors' performance. Finally, the results could be sensitive to the classification schemes used for publications and professors.

Such limitations should induce caution in interpreting the findings arising from scientometric techniques, yet we do not expect that any would be systemic to a specific Italian university.